# A WEAKLY INFORMATIVE DEFAULT PRIOR DISTRIBUTION FOR LOGISTIC AND OTHER REGRESSION MODELS

By Andrew Gelman, Aleks Jakulin, Maria Grazia Pittau and Yu-Sung Su

*Columbia University, Columbia University, University of Rome, and City University of New York*

We propose a new prior distribution for classical (nonhierarchical) logistic regression models, constructed by first scaling all nonbinary variables to have mean 0 and standard deviation 0.5, and then placing independent Student-$t$ prior distributions on the coefficients. As a default choice, we recommend the Cauchy distribution with center 0 and scale 2.5, which in the simplest setting is a longer-tailed version of the distribution attained by assuming one-half additional success and one-half additional failure in a logistic regression. Cross-validation on a corpus of datasets shows the Cauchy class of prior distributions to outperform existing implementations of Gaussian and Laplace priors.

We recommend this prior distribution as a default choice for routine applied use. It has the advantage of always giving answers, even when there is complete separation in logistic regression (a common problem, even when the sample size is large and the number of predictors is small), and also automatically applying more shrinkage to higher-order interactions. This can be useful in routine data analysis as well as in automated procedures such as chained equations for missing-data imputation.

We implement a procedure to fit generalized linear models in R with the Student-$t$ prior distribution by incorporating an approximate EM algorithm into the usual iteratively weighted least squares. We illustrate with several applications, including a series of logistic regressions predicting voting preferences, a small bioassay experiment, and an imputation model for a public health data set.

## 1. Introduction.

1.1. *Separation and sparsity in applied logistic regression.* Nonidentifiability is a common problem in logistic regression. In addition to the problem of collinearity, familiar from linear regression, discrete-data regression can also become unstable from *separation*, which arises when a linear combination of the predictors is perfectly predictive of the outcome [Albert and Anderson (1984), Lesaffre and Albert (1989)]. Separation is surprisingly common in applied logistic regression, especially with binary predictors, and, as noted by Zorn (2005), is often handled inappropriately. For example, a common "solution" to separation is to remove predictors until the resulting model is identifiable, but, as Zorn (2005) points out, this typically results in removing the strongest predictors from the model.

An alternative approach to obtaining stable logistic regression coefficients is to use Bayesian inference. Various prior distributions have been suggested for this purpose, most notably a Jeffreys prior distribution [Firth (1993)], but these have not been set up for reliable computation and are not always clearly interpretable as prior information in a regression context. Here we propose a new, proper prior distribution that produces stable, regularized estimates while still being vague enough to be used as a default in routine applied work. Our procedure can be seen as a generalization of the scaled prior distribution of Raftery (1996) to the $t$ case, with the additional innovation that the prior scale parameter is given a direct interpretation in terms of logistic regression parameters.

A simple adaptation of the usual iteratively weighted least squares algorithm allows us to estimate coefficients using independent $t$ prior distributions. This implementation works by adding pseudo-data at the least squares step and ensures numerical stability of the algorithm—in contrast to existing implementations of the Jeffreys prior distribution which can crash when applied to sparse data.

We demonstrate the effectiveness of our method in three applications: (1) a model predicting voting from demographic predictors, which is typical of many of our everyday data analyses in political science; (2) a simple bioassay model from an early article [Racine et al. (1986)] on routine applied Bayesian inference; and (3) a missing-data imputation problem from our current applied work on a study of HIV virus load. None of these applications is technically sophisticated; rather, they demonstrate the wide relevance of a default logistic regression procedure.

1.2. *Relation to existing approaches.* Our key idea is to use minimal prior knowledge, specifically that a typical change in an input variable would be unlikely to correspond to a change as large as 5 on the logistic scale (which would move the probability from 0.01 to 0.50 or from 0.50 to 0.99).

This is related to the "conditional means" approach of Bedrick, Christensen, and Johnson (1996) of setting a prior distribution by eliciting the possible distribution of outcomes given different combinations of regression inputs, and the method of Witte, Greenland, and Kim (1998) and Greenland (2001) of assigning prior distributions by characterizing expected effects in weakly informative ranges ("probably near null," "probably moderately positive," etc.). Our method differs from these related approaches in using a generic prior constraint rather than information specific to a particular analysis. As such, we would expect our prior distribution to be more appropriate for automatic use, with these other methods suggesting ways to add more targeted prior information when necessary. For example, the conditional means prior is easy to assess and the posterior is easy to fit, but it is not set up to be applied automatically to a dataset in the way that Jeffreys' prior— or ours—can be implemented. One approach for going further, discussed by MacLehose et al. (2006) and Dunson, Herring, and Engel (2006), is to use mixture prior distributions for logistic regressions with large numbers of predictors. These models use batching in the parameters, or attempt to discover such batching, in order to identify more important predictors and shrink others.

Another area of related work is the choice of parametric family for the prior distribution. We have chosen the $t$ family, focusing on the Cauchy as a conservative choice. Genkin, Lewis, and Madigan (2007) consider the Laplace (double-exponential) distribution, which has the property that its posterior mode estimates can be shrunk all the way to zero. This is an appropriate goal in projects such as text categorization (the application in that article) in which data storage is an issue, but is less relevant in social science analysis of data that have already been collected.

In the other direction, our approach (which, in the simplest logistic regression that includes only a constant term, turns out to be close to adding one-half success and one-half failure, as we discuss in Section 2.2) can be seen as a generalization of the work of Agresti and Coull (1998) on using Bayesian techniques to get point estimates and confidence intervals with good small-sample frequency properties. As we have noted earlier, similar penalized likelihood methods using the Jeffreys prior have been proposed and evaluated by Firth (1993), Heinze and Schemper (2003), Zorn (2005), and Heinze (2006). Our approach is similar but is parameterized in terms of the coefficients and thus allows us to make use of prior knowledge on that scale. In simple cases the two methods can give similar results (identical to the first decimal place in the example in Figure 3), with our algorithm being more stable by taking advantage of the existing iteratively weighted least squares algorithm.

We justify our choice of model and parameters in three ways. First, we interpret our prior distribution directly as a constraint on the logistic regression coefficients. Second, we show that our default procedure gives reasonable results in three disparate applications. Third, we borrow an idea from computer science and use cross-validation on an existing corpus of datasets to compare the predictive performance of a variety of prior distributions. The cross-validation points up the necessity of choosing between the goal of optimal predictions and the statistical principle of conservatism.

**2. A default prior specification for logistic regression.** There is a vast literature on noninformative, default, and reference prior distributions; see, Jeffreys (1961), Hartigan (1964), Bernardo (1979), Spiegelhalter and Smith (1982), Yang and Berger (1994), and Kass and Wasserman (1996). Our approach differs from most of this work in that we want to include some actual prior information, enough to regularize the extreme inferences that are obtained using maximum likelihood or completely noninformative priors. The existing literature [including, we must admit, Gelman et al. (2003)] offers the extremes of (a) fully informative prior distributions using application-specific information, or (b) noninformative priors, typically motivated by invariance principles. Our goal here is something in between: a somewhat informative prior distribution that can nonetheless be used in a wide range of applications. As always with default models, our prior can be viewed as a starting point or placeholder—a baseline on top of which the user can add real prior information as necessary. For this purpose, we want something better than the unstable estimates produced by the current default—maximum likelihood (or Bayesian estimation with a flat prior).

On the one hand, scale-free prior distributions such as Jeffreys' do not include enough prior information; on the other, what prior information can be assumed for a generic model? Our key idea is that actual effects tend to fall within a limited range. For logistic regression, a change of 5 moves a probability from 0.01 to 0.5, or from 0.5 to 0.99. We rarely encounter situations where a shift in input $x$ corresponds to the probability of outcome $y$ changing from 0.01 to 0.99, hence, we are willing to assign a prior distribution that assigns low probabilities to changes of 10 on the logistic scale.

2.1. *Standardizing input variables to a commonly-interpretable scale.* A challenge in setting up any default prior distribution is getting the scale right: for example, suppose we are predicting vote preference given age (in years). We would not want the same prior distribution if the age scale were shifted to months. But discrete predictors have their own natural scale (most notably, a change of 1 in a binary predictor) that we would like to respect.

The first step of our model is to standardize the input variables, a procedure that has been applied to Bayesian generalized linear models by Raftery (1996) and that we have formalized as follows [Gelman (2008)]:

- Binary inputs are shifted to have a mean of 0 and to differ by 1 in their lower and upper conditions. (For example, if a population is 10% African-American and 90% other, we would define the centered "African-American" variable to take on the values 0.9 and $-0.1$.)
- Other inputs are shifted to have a mean of 0 and scaled to have a standard deviation of 0.5. This scaling puts continuous variables on the same scale as symmetric binary inputs (which, taking on the values $\pm 0.5$, have standard deviation 0.5).

Following Gelman and Pardoe (2007), we distinguish between regression *inputs* and *predictors*. For example, in a regression on age, sex, and their interaction, there are four predictors (the constant term, age, sex, and age × sex), but just two inputs: age and sex. It is the input variables, not the predictors, that are standardized.

A prior distribution on standardized variables depends on the data, but this is not necessarily a bad idea. As pointed out by Raftery (1996), the data, or "the broad possible range of the variables," are relevant to knowledge about the coefficients. If we do not standardize at all, we have to worry about coefficients of very large or very small variables (for example, distance measured in millimeters, meters, or kilometers). One might follow Greenland, Schlesselman, and Criqui (2002) and require of users that they put each variable on a reasonable scale before fitting a model. Realistically, though, users routinely fit regressions on unprocessed data, and we want our default procedure to perform reasonably in such settings.

2.2. *A weakly informative t family of prior distributions.* The second step of the model is to define prior distributions for the coefficients of the predictors. We follow Raftery (1996) and assume prior independence of the coefficients as a default assumption, with the understanding that the model could be reparameterized if there are places where prior correlation is appropriate. For each coefficient, we assume a Student-$t$ prior distribution with mean 0, degrees-of-freedom parameter $\nu$, and scale $s$, with $\nu$ and $s$ chosen to provide minimal prior information to constrain the coefficients to lie in a reasonable range. We are motivated to consider the $t$ family because flat-tailed distributions allow for robust inference [see, Berger and Berliner (1986), Lange, Little, and Taylor (1989)], and, as we shall see in Section 3, it allows easy and stable computation in logistic regression by placing iteratively weighted least squares within an approximate EM algorithm. Computation with a normal prior distribution is even easier (no EM algorithm is needed), but we prefer the flexibility of the $t$ family.

Before discussing our choice of parameters, we briefly discuss some limiting cases. Setting the scale $s$ to infinity corresponds to a flat prior distribution (so that the posterior mode is the maximum likelihood estimate). As we illustrate in Section 4.1, the flat prior fails in the case of separation. Setting the degrees of freedom $\nu$ to infinity corresponds to the Gaussian distribution. As we illustrate in Section 5, we obtain better average performance by using a $t$ with finite degrees of freedom (see Figure 6).[1] We suspect that the Cauchy prior distribution outperforms the normal, on average, because it allows for occasional large coefficients while still performing a reasonable amount of shrinkage for coefficients near zero; this is another way of saying that we think the set of true coefficients that we might encounter in our logistic regressions has a distribution less like a normal than like a Cauchy, with many small values and occasional large ones.

One way to pick a default value of $\nu$ and $s$ is to consider the baseline case of one-half of a success and one-half of a failure for a single binomial trial with probability $p = \text{logit}^{-1}(\theta)$—that is, a logistic regression with only a constant term. The corresponding likelihood is $e^{\theta/2}/(1 + e^\theta)$, which is close to a $t$ density function with 7 degrees of freedom and scale 2.5 [Liu (2004)]. We shall choose a slightly more conservative choice, the Cauchy, or $t_1$, distribution, again with a scale of 2.5. Figure 1 shows the three density functions: they all give preference to values less than 5, with the Cauchy allowing the occasional possibility of very large values (a point to which we return in Section 5).

We assign independent Cauchy prior distributions with center 0 and scale 2.5 to each of the coefficients in the logistic regression except the constant term. When combined with the standardization, this implies that the absolute difference in logit probability should be less then 5, when moving from one standard deviation below the mean, to one standard deviation above the mean, in any input variable.

If we were to apply this prior distribution to the constant term as well, we would be stating that the success probability is probably between 1% and 99% for units that are average in all the inputs. Depending on the context [for example, epidemiologic modeling of rare conditions, as in Greenland (2001)], this might not make sense, so as a default we apply a weaker prior

---

[1]In his discussion of default prior distributions for generalized linear models, Raftery (1996) works with the Gaussian family and writes that "the results depend little on the precise functional form." One reason that our recommendations differ in their details from Raftery's is that we are interested in predictions and inferences within a single model, with a particular interest in sparse data settings where the choice of prior distribution can make a difference. In contrast, Raftery's primary interest in his 1996 paper lay in the effect of the prior distribution on the marginal likelihood and its implications for the Bayes factor as used in model averaging.

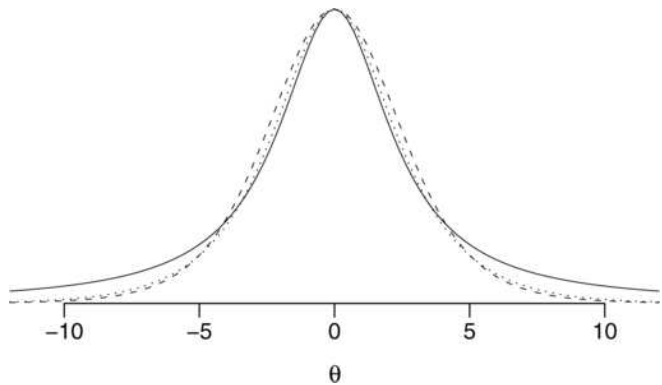

FIG. 1. *(Solid line) Cauchy density function with scale 2.5, (dashed line) $t_7$ density function with scale 2.5, (dotted line) likelihood for $\theta$ corresponding to a single binomial trial of probability $\text{logit}^{-1}(\theta)$ with one-half success and one-half failure. All these curves favor values below 5 in absolute value; we choose the Cauchy as our default model because it allows the occasional probability of larger values.*

distribution—a Cauchy with center 0 and scale 10, which implies that we expect the success probability for an average case to be between $10^{-9}$ and $1 - 10^{-9}$.

An appealing byproduct of applying the model to rescaled predictors is that it automatically implies more stringent restrictions on interactions. For example, consider three symmetric binary inputs, $x_1, x_2, x_3$. From the rescaling, each will take on the values $\pm 1/2$. Then any two-way interaction will take on the values $\pm 1/4$, and the three-way interaction can be $\pm 1/8$. But all these coefficients have the same default prior distribution, so the total contribution of the three-way interaction is $1/4$ that of the main effect. Going from the low value to the high value in any given three-way interaction is, in the model, unlikely to change the logit probability by more than $5 \cdot (1/8 - (-1/8)) = 5/4$ on the logit scale.

**3. Computation.** In principle, logistic regression with our prior distribution can be computed using the Gibbs and Metropolis algorithms. We do not give details as this is now standard with Bayesian models; see, for example, Carlin and Louis (2001), Martin and Quinn (2002), and Gelman et al. (2003). In practice, however, it is desirable to have a quick calculation that returns a point estimate of the regression coefficients and standard errors. Such an approximate calculation works in routine statistical practice and, in addition, recognizes the approximate nature of the model itself.

We consider three computational settings:

- Classical (nonhierarchical) logistic regression, using our default prior distribution in place of the usual flat prior distribution on the coefficients.

- Multilevel (hierarchical) modeling, in which some of the default prior distribution is applied only to the subset of the coefficients that are not otherwise modeled (sometimes called the "fixed effects").
- Chained imputation, in which each variable with missing data is modeled conditional on the other variables with a regression equation, and these models are fit and random imputations inserted iteratively [Van Buuren and Oudshoom (2000), Raghunathan, Van Hoewyk, and Solenberger (2001)].

In any of these cases, our default prior distribution has the purpose of stabilizing (regularizing) the estimates of otherwise unmodeled parameters. In the first scenario, the user typically only extracts point estimates and standard errors. In the second scenario, it makes sense to embed the computation within the full Markov chain simulation. In the third scenario of missing-data imputation, we would like the flexibility of quick estimates for simple problems with the potential for Markov chain simulation as necessary. Also, because of the automatic way in which the component models are fit in a chained imputation, we would like a computationally stable algorithm that returns reasonable answers.

We have implemented these computations by altering the `glm` function in R, creating a new function, `bayesglm`, that finds an approximate posterior mode and variance using extensions of the classical generalized linear model computations, as described in the rest of this section. The `bayesglm` function (part of the `arm` package for applied regression and multilevel modeling in R) allows the user to specify independent prior distributions for the coefficients in the $t$ family, with the default being Cauchy distributions with center 0 and scale set to 10 (for the regression intercept), 2.5 (for binary predictors), or $2.5/(2 \cdot \text{sd})$, where `sd` is the standard deviation of the predictor in the data (for other numerical predictors). We are also extending the program to fit hierarchical models in which regression coefficients are structured in batches [Gelman et al. (2008)].

3.1. *Incorporating the prior distribution into classical logistic regression computations.* Working in the context of the logistic regression model,

$$\Pr(y_i = 1) = \text{logit}^{-1}(X_i \beta), \tag{1}$$

we adapt the classical maximum likelihood algorithm to obtain approximate posterior inference for the coefficients $\beta$, in the form of an estimate $\hat{\beta}$ and covariance matrix $V_\beta$.

The standard logistic regression algorithm—upon which we build—proceeds by approximately linearizing the derivative of the log-likelihood, solving using weighted least squares, and then iterating this process, each step evaluating the derivatives at the latest estimate $\hat{\beta}$; see, for example, McCullagh and Nelder (1989). At each iteration, the algorithm determines pseudo-data $z_i$ and pseudo-variances $(\sigma_i^z)^2$ based on the linearization of the derivative of the log-likelihood,

$$z_i = X_i\hat{\beta} + \frac{(1+e^{X_i\hat{\beta}})^2}{e^{X_i\hat{\beta}}}\left(y_i - \frac{e^{X_i\hat{\beta}}}{1+e^{X_i\hat{\beta}}}\right),$$

(2) $$(\sigma_i^z)^2 = \frac{1}{n_i}\frac{(1+e^{X_i\hat{\beta}})^2}{e^{X_i\hat{\beta}}},$$

and then performs weighted least squares, regressing $z$ on $X$ with weight vector $(\sigma^z)^{-2}$. The resulting estimate $\hat{\beta}$ is used to update the computations in (2), and the iteration proceeds until approximate convergence.

*Computation with a specified normal prior distribution.* The simplest informative prior distribution assigns normal prior distributions for the components of $\beta$:

(3) $$\beta_j \sim N(\mu_j, \sigma_j^2) \qquad \text{for } j=1,\ldots,J.$$

This information can be effortlessly included in the classical algorithm by simply altering the weighted least-squares step, augmenting the approximate likelihood with the prior distribution; see, for example, Section 14.8 of Gelman et al. (2003). If the model has $J$ coefficients $\beta_j$ with independent $N(\mu_j, \sigma_j^2)$ prior distributions, then we add $J$ pseudo-data points and perform weighted linear regression on "observations" $z_*$, "explanatory variables" $X_*$, and weight vector $w_*$, where

(4) $$z_* = \begin{pmatrix} z \\ \mu \end{pmatrix}, \qquad X_* = \begin{pmatrix} X \\ I_J \end{pmatrix}, \qquad w_* = (\sigma^z, \sigma)^{-2}.$$

The vectors $z_*, w_*$, and the matrix $X_*$ are constructed by combining the likelihood [$z$ and $\sigma^z$, are the vectors of $z_i$'s and $\sigma_i^z$'s defined in (2), and $X$ is the design matrix of the regression (1)] and the prior [$\mu$ and $\sigma$ are the vectors of $\mu_j$'s and $\sigma_j$'s in (3), and $I_J$ is the $J \times J$ identity matrix]. As a result, $z_*$ and $w_*$ are vectors of length $n+J$ and $X_*$ is an $(n+J) \times J$ matrix. With the augmented $X_*$, this regression is identified, and, thus, the resulting estimate $\hat{\beta}$ is well defined and has finite variance, even if the original data have collinearity or separation that would result in nonidentifiability of the maximum likelihood estimate.

The full computation is then iteratively weighted least squares, starting with a guess of $\beta$ (for example, independent draws from the unit normal distribution), then computing the derivatives of the log-likelihood to compute $z$ and $\sigma_z$, then using weighted least squares on the pseudo-data (4) to yield an updated estimate of $\beta$, then recomputing the derivatives of the log-likelihood at this new value of $\beta$, and so forth, converging to the estimate $\hat{\beta}$.

The covariance matrix $V_\beta$ is simply the inverse second derivative matrix of the log-posterior density evaluated at $\hat{\beta}$—that is, the usual normal-theory uncertainty estimate for an estimate not on the boundary of parameter space.

*Approximate EM algorithm with a t prior distribution.* If the coefficients $\beta_j$ have independent $t$ prior distributions[2] with centers $\mu_j$ and scales $s_j$, we can adapt the just-described iteratively weighted least squares algorithm to estimate the coefficients using an approximate EM algorithm (Dempster, Laird and Rubin 1977). We shall describe the steps of the algorithm shortly; the idea is to express the $t$ prior distribution for each coefficient $\beta_j$ as a mixture of normals with unknown scale $\sigma_j$:

$$\text{(5)} \qquad \beta_j \sim \text{N}(\mu_j, \sigma_j^2), \qquad \sigma_j^2 \sim \text{Inv-}\chi^2(\nu_j, s_j^2)$$

and then average over the $\beta_j$'s at each step, treating them as missing data and performing the EM algorithm to estimate the $\sigma_j$'s. The algorithm proceeds by alternating one step of iteratively weighted least squares (as described above) and one step of EM. Once enough iterations have been performed to reach approximate convergence, we get an estimate and covariance matrix for the vector parameter $\beta$ and the estimated $\sigma_j$'s.

We initialize the algorithm by setting each $\sigma_j$ to the value $s_j$ (the scale of the prior distribution) and, as before, starting with a guess of $\beta$ (either obtained from a simpler procedure or simply picking a starting value such as $\beta = 0$). Then, at each step of the algorithm, we update $\sigma$ by maximizing the expected value of its (approximate) log-posterior density,

$$\log p(\beta, \sigma | y) \approx -\frac{1}{2} \sum_{i=1}^{n} \frac{1}{(\sigma_i^z)^2} (z_i - X_i \beta)^2$$

$$-\frac{1}{2} \sum_{j=1}^{J} \left( \frac{1}{\sigma_j^2} (\beta_j - \mu_j)^2 + \log(\sigma_j^2) \right)$$

$$\text{(6)} \qquad - p(\sigma_j | \nu_j, s_j) + \text{constant}.$$

Each iteration of the algorithm proceeds as follows:

1. Based on the current estimate of $\beta$, perform the normal approximation to the log-likelihood and determine the vectors $z$ and $\sigma^z$ using (2), as in classical logistic regression computation.

---

[2] As discussed earlier, we use the default settings $\mu_j = 0$, $s_j = 2.5$, $\nu_j = 1$ (except for the constant term, if any, to whose prior distributions we assign the parameters $\mu_j = 0$, $s_j = 10$, $\nu_j = 1$), but we describe the computation more generally in terms of arbitrary values of these parameters.

2. Approximate E-step: first run the weighted least squares regression based on the augmented data (4) to get an estimate $\hat{\beta}$ with variance matrix $V_\beta$. Then determine the expected value of the log-posterior density by replacing the terms $(\beta_j - \mu_j)^2$ in (6) by

(7) $$\mathrm{E}((\beta_j - \mu_j)^2 | \sigma, y) \approx (\hat{\beta}_j - \mu_j)^2 + (V_\beta)_{jj},$$

which is only approximate because we are averaging over a normal distribution that is only an approximation to the generalized linear model likelihood.

3. M-step: maximize the (approximate) expected value of the log-posterior density (6) to get the estimate,

(8) $$\hat{\sigma}_j^2 = \frac{(\hat{\beta}_j - \mu_j)^2 + (V_\beta)_{jj} + \nu_j s_j^2}{1 + \nu_j},$$

which corresponds to the (approximate) posterior mode of $\sigma_j^2$ given a single measurement with value (7) and an Inv-$\chi^2(\nu_j, s_j^2)$ prior distribution.

4. Recompute the derivatives of the log-posterior density given the current $\hat{\beta}$, set up the augmented data (4) using the estimated $\hat{\sigma}$ from (8), and repeat steps 1, 2, 3 above.

At convergence of the algorithm, we summarize the inferences using the latest estimate $\hat{\beta}$ and covariance matrix $V_\beta$.

3.2. *Other models.*

*Linear regression.* Our algorithm is basically the same for linear regression, except that weighted least squares is an exact rather than approximate maximum penalized likelihood, and also a step needs to be added to estimate the data variance. In addition, we would preprocess $y$ by rescaling the outcome variable to have mean 0 and standard deviation 0.5 before assigning the prior distribution (or, equivalently, multiply the prior scale parameter by the standard deviation of the data). Separation is not a concern in linear regression; however, when applied routinely (for example, in iterative imputation algorithms), collinearity can arise, in which case it is helpful to have a proper but weak prior distribution.

*Other generalized linear models.* Again, the basic algorithm is unchanged, except that the pseudo-data and pseudo-variances in (2), which are derived from the first and second derivatives of the log-likelihood, are changed [see Section 16.4 of Gelman et al. (2003)]. For Poisson regression and other models with the logarithmic link, we would not often expect effects larger than 5 on the logarithmic scale, and so the prior distributions given in this article might be a reasonable default choice. In addition, for models such as the negative binomial that have dispersion parameters, these can be estimated using an additional step as is done when estimating the data-level variance in normal linear regression. For more complex models such as multinomial logit and probit, we have considered combining independent t prior distributions on the coefficients with pseudo-data to identify cutpoints in the possible presence of sparse data. Such models also present computational challenges, as there is no simple existing iteratively weighted least squares algorithm for us to adapt.

*Avoiding nested looping when inserting into larger models.* In multilevel models [Gelman et al. (2008)] or in applications such as chained imputation (discussed in Section 4.3), it should be possible to speed the computation by threading, rather than nesting, the loops. For example, suppose we are fitting an imputation by iteratively regressing $u$ on $v, w$, then $v$ on $u, w$, then $w$ on $u, v$. Instead of doing a full iterative weighted least squares at each iteration, then we could perform one step of weighted least squares at each step, thus taking less computer time to ultimately converge by not wasting time by getting hyper-precise estimates at each step of the stochastic algorithm.

## 4. Applications.

4.1. *A series of regressions predicting vote preferences.* Regular users of logistic regression know that separation can occur in routine data analyses, even when the sample size is large and the number of predictors is small. The left column of Figure 2 shows the estimated coefficients for logistic regression predicting the probability of a Republican vote for president for a series of elections. The estimates look fine except in 1964, where there is complete separation, with all the African-American respondents supporting the Democrats. Fitting in R actually yields finite estimates, as displayed in the graph, but these are essentially meaningless, being a function of how long the iterative fitting procedure goes before giving up.

The other three columns of Figure 2 show the coefficient estimates using our default Cauchy prior distribution for the coefficients, along with the $t_7$ and normal distributions. (In all cases, the prior distributions are centered at 0, with scale parameters set to 10 for the constant term and 2.5 for all other coefficients.) All three prior distributions do a reasonable job at stabilizing the estimated coefficient for race for 1964, while leaving the estimates for other years essentially unchanged. This example illustrates how we could use our Bayesian procedure in routine practice.

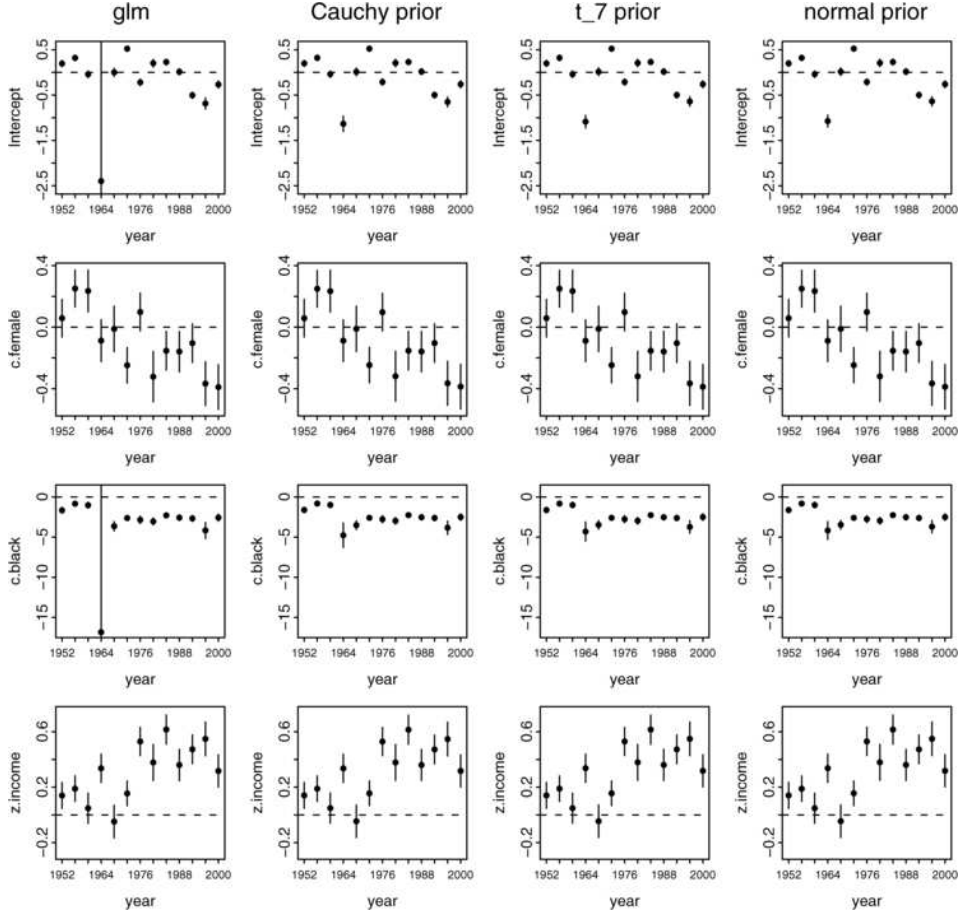

Fig. 2. *The left column shows the estimated coefficients ($\pm 1$ standard error) for a logistic regression predicting the probability of a Republican vote for president given sex, race, and income, as fit separately to data from the National Election Study for each election 1952 through 2000. [The binary inputs* female *and* black *have been centered to have means of zero, and the numerical variable* income *(originally on a 1–5 scale) has been centered and then rescaled by dividing by two standard deviations.]*

*There is complete separation in 1964 (with none of the black respondents supporting the Republican candidate, Barry Goldwater), leading to a coefficient estimate of $-\infty$ that year. (The particular finite values of the estimate and standard error are determined by the number of iterations used by the* glm *function in R before stopping.)*

*The other columns show estimated coefficients ($\pm 1$ standard error) for the same model fit each year using independent Cauchy, $t_7$, and normal prior distributions, each with center 0 and scale 2.5. All three prior distributions do a reasonable job at stabilizing the estimates for 1964, while leaving the estimates for other years essentially unchanged.*

4.2. *A small bioassay experiment.* We next consider a small-sample example in which the prior distribution makes a difference for a coefficient

14 GELMAN, JAKULIN, PITTAU AND SU

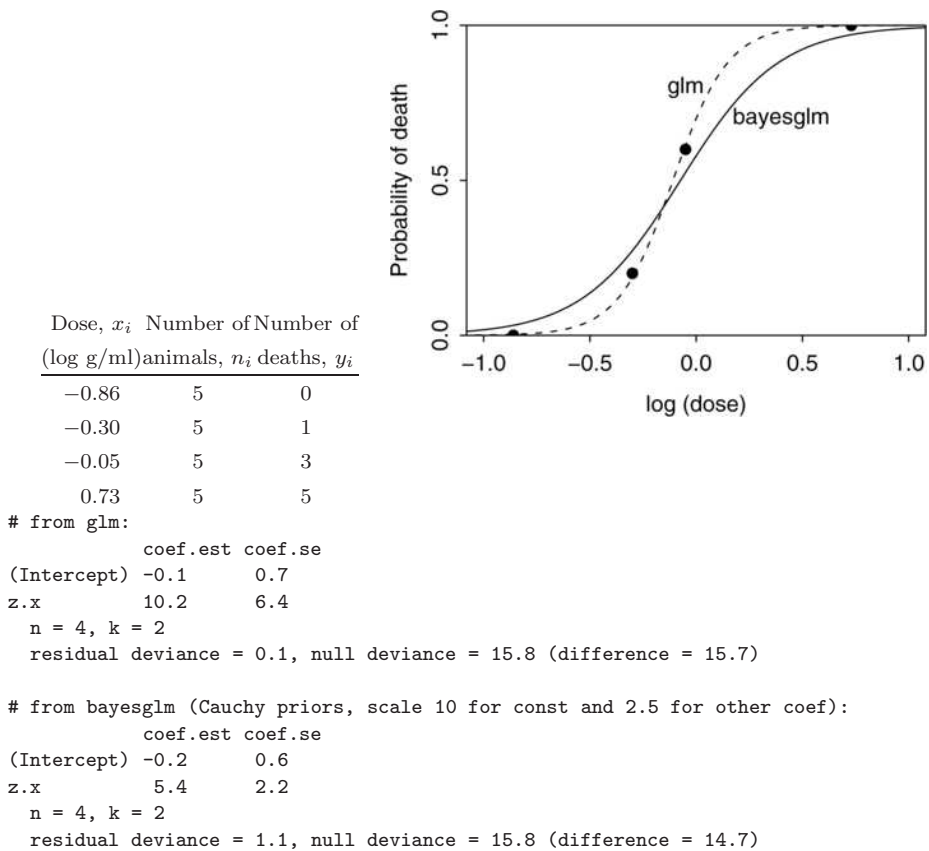

| Dose, $x_i$ (log g/ml) | Number of animals, $n_i$ | Number of deaths, $y_i$ |
|---|---|---|
| $-0.86$ | 5 | 0 |
| $-0.30$ | 5 | 1 |
| $-0.05$ | 5 | 3 |
| 0.73 | 5 | 5 |

```
# from glm:
            coef.est coef.se
(Intercept) -0.1     0.7
z.x          10.2    6.4
  n = 4, k = 2
  residual deviance = 0.1, null deviance = 15.8 (difference = 15.7)

# from bayesglm (Cauchy priors, scale 10 for const and 2.5 for other coef):
            coef.est coef.se
(Intercept) -0.2     0.6
z.x          5.4     2.2
  n = 4, k = 2
  residual deviance = 1.1, null deviance = 15.8 (difference = 14.7)
```

FIG. 3. *Data from a bioassay experiment, from Racine et al. (1986), and estimates from classical maximum likelihood and Bayesian logistic regression with the recommended default prior distribution. In addition to graphing the fitted curves (at top right), we show raw computer output to illustrate how our approach would be used in routine practice.*
*The big change in the estimated coefficient for* `z.x` *when going from* `glm` *to* `bayesglm` *may seem surprising at first, but upon reflection we prefer the second estimate with its lower coefficient for x, which is based on downweighting the most extreme possibilities that are allowed by the likelihood.*

that is already identified. The example comes from Racine et al. (1986), who used a problem in bioassay to illustrate how Bayesian inference can be applied with small samples. The top part of Figure 3 presents the data, from twenty animals that were exposed to four different doses of a toxin. The bottom parts of Figure 3 show the resulting logistic regression, as fit first using maximum likelihood and then using our default Cauchy prior distributions with center 0 and scale 10 (for the constant term) and 2.5 (for the coefficient of dose). Following our general procedure, we have rescaled dose to have mean 0 and standard deviation 0.5.

With such a small sample, the prior distribution actually makes a difference, lowering the estimated coefficient of standardized dose from $10.2 \pm 6.4$ to $5.4 \pm 2.2$. Such a large change might seem disturbing, but for the reasons discussed above, we would doubt the effect to be as large as 10.2 on the logistic scale, and the analysis shows these data to be consistent with the much smaller effect size of 5.4. The large amount of shrinkage simply confirms how weak the information is that gave the original maximum likelihood estimate. The graph at the upper right of Figure 3 shows the comparison in a different way: the maximum likelihood estimate fits the data almost perfectly; however, the discrepancies between the data and the Bayes fit are small, considering the sample size of only 5 animals within each group.[3]

4.3. *A set of chained regressions for missing-data imputation.* Multiple imputation [Rubin (1978, 1996)] is another context in which regressions with many predictors are fit in an automatic way. It is common to have missing data in several variables in an analysis, in which case one cannot simply set up a model for a single partially-observed outcome given a set of fully-observed predictors. More generally, we must think of the dataset as a multivariate outcome, any components of which can be missing. The direct approach to imputing missing data in several variables is to fit a multivariate model. However, this approach requires a lot of effort to set up a reasonable multivariate regression model and a fully specified joint model is sometime difficult to specify, particularly when we have a mixture of different types of variables.

A different approach, becoming more popular for imputing missing data, uses chained equations [Van Buuren and Oudshoom (2000), Raghunathan, Van Hoewyk, and Solenberger (2001)], a series of conditional distributions without the need to fit a multivariate model. In chained imputation, each variable is imputed using a regression model conditional on all the others, iteratively cycling through all the variables that contain missing data. Different models can be specified for different variables to be imputed, and logistic regression is a natural choice for binary variables. When the number of variables is large, separation can arise. Our prior distribution yields stable computations in this setting, as we illustrate in an example from our current applied research.

We consider a model from our current applied research imputing virus loads in a longitudinal sample of HIV-positive homeless persons. The analysis incorporates a large number of predictors, including demographic and

---

[3]For example, the second data point $(\log(x) = -0.30)$ has an empirical rate of $1/5 = 0.20$ and a predicted probability (from the Bayes fit) of 0.27. With a sample size of 5, we could expect a standard error of $\sqrt{0.27 \cdot (1 - 0.27)/5} = 0.20$, so a difference of 0.07 should be of no concern.

```
# from glm:
                 coef.est coef.sd                   coef.est coef.sd
(Intercept)          0.07    1.41   h39b.W1            -0.10 0.03
age.W1               0.02    0.02   pcs.W1             -0.01 0.01
mcs37.W1            -0.01    0.32   nonhaartcombo.W1  -20.99 888.74
unstabl.W1          -0.09    0.37   b05.W1             -0.07 0.12
ethnic.W3           -0.14    0.23   h39b.W2             0.02 0.03
age.W2               0.02    0.02   pcs.W2             -0.01 0.02
mcs37.W2             0.26    0.31   haart.W2            1.80 0.30
nonhaartcombo.W2     1.33    0.44   unstabl.W2          0.27 0.42
b05.W2               0.03    0.12   h39b.W3             0.00 0.03
age.W3              -0.01    0.02   pcs.W3              0.01 0.01
mcs37.W3            -0.04    0.32   haart.W3            0.60 0.31
nonhaartcombo.W3     0.44    0.42   unstabl.W3         -0.92 0.40
b05.W3              -0.11    0.11

# from bayesglm (Cauchy priors, scale 10 for const
                       and 2.5 for other coefs):
                 coef.est coef.sd                   coef.est coef.sd
(Intercept)         -0.84    1.15   h39b.W1            -0.08 0.03
age.W1               0.01    0.02   pcs.W1             -0.01 0.01
mcs37.W1            -0.10    0.31   nonhaartcombo.W1   -6.74 1.22
unstabl.W1          -0.06    0.36   b05.W1              0.02 0.12
ethnic.W3            0.18    0.21   h39b.W2             0.01 0.03
age.W2               0.03    0.02   pcs.W2             -0.02 0.02
mcs37.W2             0.19    0.31   haart.W2            1.50 0.29
nonhaartcombo.W2     0.81    0.42   unstabl.W2          0.29 0.41
b05.W2               0.11    0.12   h39b.W3            -0.01 0.03
age.W3              -0.02    0.02   pcs.W3              0.01 0.01
mcs37.W3             0.05    0.32   haart.W3            1.02 0.29
nonhaartcombo.W3     0.64    0.40   unstabl.W3         -0.52 0.39
b05.W3              -0.15    0.13
```

Fig. 4. *A logistic regression fit for missing-data imputation using maximum likelihood (top) and Bayesian inference with default prior distribution (bottom). The classical fit resulted in an error message indicating separation; in contrast, the Bayes fit (using independent Cauchy prior distributions with mean 0 and scale 10 for the intercept and 2.5 for the other coefficients) produced stable estimates. We would not usually summarize results using this sort of table, however, this gives a sense of how the fitted models look in routine data analysis.*

health-related variables, and often with high rates of missingness. Inside the multiple imputation chained equation procedure, logistic regression is used to impute the binary variables. It is generally recommended to include a rich set of predictors when imputing missing values [Rubin (1996)]. However, in this application, including all the dichotomous predictors leads to many instances of separation.

To take one example from our analysis, separation arose when estimating each person's probability of attendance in a group therapy called haart. The top part of Figure 4 shows the model as estimated using the glm function in R fit to the observed cases in the first year of the data set: the coefficient for nonhaartcombo.W1 is essentially infinity, and the regression also gives an error message indicating nonidentifiability. The bottom part of Figure 4 shows the fit using our recommended Bayesian procedure (this time, for simplicity, not recentering and rescaling the inputs, most of which are actually binary).

In the chained imputation, the classical glm fits were nonidentifiable at many places; none of these presented any problem when we switched to our new bayesglm function.[4]

**5. Data from a large number of logistic regressions.** In the spirit of Stigler (1977), we wanted to see how large are logistic regression coefficients in some general population, to get a rough sense of what would be a reasonable default prior distribution. One way to do this is to fit many logistic regressions to available data sets and estimate the underlying distribution of coefficients. Another approach, which we follow here, is to examine the cross-validated predictive quality of different types of priors on a corpus of data sets, following the approach of meta-learning in computer science; see, for example, Vilalta and Drissi (2002).

5.1. *Cross-validation on a corpus of data sets.* The fundamental idea of predictive modeling is that the data are split into two subsets, the training and the test data. The training data are used to construct a model, and the performance of the model on the test data is used to check whether the predictions generalize well. Cross-validation is a way of creating several different partitions. For example, assume that we put aside 1/5 of the data for testing. We divide up the data into 5 pieces of the same size. This creates 5 different partitions, and for each experiment we take one of the pieces as the test set and all the others as the training set. In the present section we summarize our efforts in evaluating our prior distribution from the predictive perspective.

For each of the random divisions of a dataset into training and test sets, our predictive evaluation takes the Bayesian point estimate fit from the training data, uses the predictors from the test set to get predicted probabilities of the 0 and 1 outcomes for each point, then compares these to the actual outcomes in the test data. We are not, strictly speaking, evaluating the prior distribution; rather, we are evaluating the point estimate (the posterior mode) derived from the specified prior. This makes sense for evaluating logistic regression methods to be used in routine practice, which typically comes down to point estimates (as in many regression summaries) or predictions (as in multiple imputation). To compare different priors for fully Bayesian inference, it might make sense to look at properties of posterior simulations, but we do not do that more computationally elaborate procedure here.

---

[4]We also tried the brlr and brglm functions in R, which implement the Jeffreys prior distributions of Firth (1993) and Kosimidis (2007). Unfortunately, we still encountered problems in achieving convergence and obtaining reasonable answers, several times obtaining an error message indicating nonconvergence of the optimization algorithm. We suspect brlr has problems because it uses a general-purpose optimization algorithm that, when fitting regression models, is less stable than iteratively weighted least squares. The brglm function uses iteratively weighted least squares and is more reliable than brlr; see Section 5.2.

| Name | Cases | Num | Cat | Pred | Outcome | $\Pr(y=1)$ | $\Pr(\text{NA})$ | $\overline{|\vec{x}|}$ |
|---|---|---|---|---|---|---|---|---|
| mushroom | 8124 | 0 | 22 | 95 | edible=e | 0.52 | 0 | 3.0 |
| spam | 4601 | 57 | 0 | 105 | class=0 | 0.61 | 0 | 3.2 |
| krkp | 3196 | 0 | 36 | 37 | result=won | 0.52 | 0 | 2.6 |
| segment | 2310 | 19 | 0 | 154 | y=5 | 0.14 | 0 | 3.5 |
| titanic | 2201 | 0 | 3 | 5 | surv=no | 0.68 | 0 | 0.7 |
| car | 1728 | 0 | 6 | 15 | eval=unacc | 0.70 | 0 | 2.0 |
| cmc | 1473 | 2 | 7 | 19 | Contracept=1 | 0.43 | 0 | 1.9 |
| german | 1000 | 7 | 13 | 48 | class=1 | 0.70 | 0 | 2.8 |
| tic-tac-toe | 958 | 0 | 9 | 18 | y=p | 0.65 | 0 | 2.3 |
| heart | 920 | 7 | 6 | 30 | num=0 | 0.45 | 0.15 | 2.3 |
| anneal | 898 | 6 | 32 | 64 | y=3 | 0.76 | 0.65 | 2.4 |
| vehicle | 846 | 18 | 0 | 58 | Y=3 | 0.26 | 0 | 3.0 |
| pima | 768 | 8 | 0 | 11 | class=0 | 0.65 | 0 | 1.8 |
| crx | 690 | 6 | 9 | 45 | A16=- | 0.56 | 0.01 | 2.3 |
| australian | 690 | 6 | 8 | 36 | Y=0 | 0.56 | 0 | 2.3 |
| soybean-large | 683 | 35 | 0 | 75 | y=brown-spot | 0.13 | 0.10 | 3.2 |
| breast-wisc-c | 683 | 9 | 0 | 20 | y=2 | 0.65 | 0 | 1.6 |
| balance-scale | 625 | 0 | 4 | 16 | name=L | 0.46 | 0 | 1.8 |
| monk2 | 601 | 0 | 6 | 11 | y=0 | 0.66 | 0 | 1.9 |
| wdbc | 569 | 20 | 0 | 45 | diag=B | 0.63 | 0 | 3.0 |
| monk1 | 556 | 0 | 6 | 11 | y=0 | 0.50 | 0 | 1.9 |
| monk3 | 554 | 0 | 6 | 11 | y=1 | 0.52 | 0 | 1.9 |
| voting | 435 | 0 | 16 | 32 | party=dem | 0.61 | 0 | 2.7 |
| horse-colic | 369 | 7 | 19 | 121 | outcom=1 | 0.61 | 0.20 | 3.4 |
| ionosphere | 351 | 32 | 0 | 110 | y=g | 0.64 | 0 | 3.5 |
| bupa | 345 | 6 | 0 | 6 | selector=2 | 0.58 | 0 | 1.5 |
| primary-tumor | 339 | 0 | 17 | 25 | primary=1 | 0.25 | 0.04 | 2.0 |
| ecoli | 336 | 7 | 0 | 12 | y=cp | 0.43 | 0 | 1.3 |
| breast-LJ-c | 286 | 3 | 6 | 16 | recurrence=no | 0.70 | 0.01 | 1.8 |
| shuttle-control | 253 | 0 | 6 | 10 | y=2 | 0.57 | 0 | 1.8 |
| audiology | 226 | 0 | 69 | 93 | y=cochlear-age | 0.25 | 0.02 | 2.3 |
| glass | 214 | 9 | 0 | 15 | y=2 | 0.36 | 0 | 1.7 |
| yeast-class | 186 | 79 | 0 | 182 | func=Ribo | 0.65 | 0.02 | 4.6 |
| wine | 178 | 13 | 0 | 24 | Y=2 | 0.40 | 0 | 2.2 |
| hayes-roth | 160 | 0 | 4 | 11 | y=1 | 0.41 | 0 | 1.5 |
| hepatitis | 155 | 6 | 13 | 35 | Class=LIVE | 0.79 | 0.06 | 2.5 |
| iris | 150 | 4 | 0 | 8 | y=virginica | 0.33 | 0 | 1.6 |
| lymphography | 148 | 2 | 16 | 29 | y=2 | 0.55 | 0 | 2.5 |
| promoters | 106 | 0 | 57 | 171 | y=mm | 0.50 | 0 | 6.1 |
| zoo | 101 | 1 | 15 | 17 | type=mammal | 0.41 | 0 | 2.2 |
| post-operative | 88 | 1 | 7 | 14 | ADM-DECS=A | 0.73 | 0.01 | 1.6 |
| soybean-small | 47 | 35 | 0 | 22 | y=D4 | 0.36 | 0 | 2.6 |
| lung-cancer | 32 | 0 | 56 | 103 | y=2 | 0.41 | 0 | 4.3 |
| lenses | 24 | 0 | 4 | 5 | lenses=none | 0.62 | 0 | 1.4 |
| o-ring-erosion | 23 | 3 | 0 | 4 | no-therm-d=0 | 0.74 | 0 | 0.7 |

FIG. 5. *The 45 datasets from the UCI Machine Learning data repository which we used for our cross-validation. Each dataset is described with its name, the number of cases in it (Cases), the number of numerical attributes (Num), the number of categorical attributes (Cat), the number of binary predictors generated from the initial set of attributes by means of discretization (Pred), the event corresponding to the positive binary outcome (Outcome), the percentage of cases having the positive outcome ($p_{y=1}$), the proportion of attribute values that were missing, expressed as a percentage (NA), and the average length of the predictor vector, ($\overline{|\vec{x}|}$).*

Performance of an estimator can be summarized in a single number for a whole data set (using expected squared error or expected log error), and so we can work with a larger collection of data sets, as is customary in machine learning. For our needs we have taken a number of data sets from the UCI Machine Learning Repository [Newman et al. (1998), Asuncion and Newman (2007)], disregarding those whose outcome is a continuous variable (such as "anonymous Microsoft Web data") and those that are given in the form of logical theories (such as "artificial characters"). Figure 5 summarized the datasets we used for our cross-validation.

Because we do not want our results to depend on an imputation method, we treat missingness as a separate category for each variable for which there are missing cases: that is, we add an additional predictor for each variable with missing data indicating whether the particular predictor's value is missing. We also use the Fayyad and Irani (1993) method for converting continuous predictors into discrete ones. To convert a $k$-level predictor into a set of binary predictors, we create $k-1$ predictors corresponding to all levels except the most frequent. Finally, for all data sets with multinomial outcomes, we transform into binary by simply comparing the most frequent category to the union of all the others.

5.2. *Average predictive errors corresponding to different prior distributions.* We use fivefold cross-validation to compare "bayesglm" (our approximate Bayes point estimate) for different default scale and degrees of freedom parameters; recall that degrees of freedom equal 1 and $\infty$ for the Cauchy and Gaussian prior distributions, respectively. We also compare to three existing methods: (1) the "glm" function in R that fits classical logistic regression (equivalent to bayesglm with prior scale set to $\infty$); (2) the "brglm" implementation of Jeffreys' prior from Kosmidis (2007), with logit and probit links; and (3) the BBR (Bayesian binary regression) algorithm of Genkin, Lewis, and Madigan (2007), which adaptively sets the scale for the choice of Laplacian or Gaussian prior distribution.

In comparing with glm, we had a practical constraint. When no finite maximum likelihood estimate exists, we define the glm solution as that obtained by the R function using its default starting value and default number of iterations.

Figure 6 shows the results, displaying average logarithmic and Brier score losses for different choices of prior distribution.[5] The Cauchy prior distribution with scale 0.75 performs best, on average. Classical logistic regression ("glm"), which corresponds to prior degrees of freedom and prior scale both set to $\infty$, did not do well: with no regularization, maximum likelihood occasionally gives extreme estimates, which then result in large penalties in the cross-validation. In fact, the log and Brier scores for classical logistic regression would be even worse except that the glm function in R stops after a finite number of iterations, thus giving estimates that are less extreme than they would otherwise be. Surprisingly, Jeffreys' prior, as implemented in brglm, also performed poorly in the cross-validation. The second-order unbiasedness property of Jeffreys' prior, while theoretically defensible [see Kosmidis (2007)], does not make use of some valuable prior information, notably that changes on the logistic scale are unlikely to be more than 5 (see Section 2.2).

The Cauchy prior distribution with scale 0.75 is a good consensus choice, but for any particular dataset, other prior distributions can perform better. To illustrate, Figure 7 shows the cross-validation errors for individual data sets in the corpus for the Cauchy prior distribution with different choices of the degrees-of-freedom and scale parameter. The Cauchy (for example, $t_1$) performs reasonably well in both cases, and much better than classical glm, but the optimal prior distribution is different for each particular dataset.

5.3. *Choosing a weakly-informative prior distribution.* The Cauchy prior distribution with scale 0.75 performs the best, yet we recommend as a default a larger scale of 2.5. Why? The argument is that, following the usual principles of noninformative or weakly informative prior distributions, we are including in our model less information than we actually have. This approach is generally considered "conservative" in statistical practice [Gelman and Jakulin (2007)]. In the case of logistic regression, the evidence suggests that the Cauchy distribution with scale 0.75 captures the underlying variation in logistic regression coefficients in a corpus of data sets. We use a scale of 2.5 to weaken this prior information and bring things closer to the

---

[5]Given the vector of predictors $\vec{x}$, the true outcome $y$ and the predicted probability $p_y = f(\vec{x})$ for $y$, the Brier score is defined as $(1-p_y)^2/2$ and the logarithmic score is defined as $-\log p_y$. Because of cross-validation, the probabilities were built without using the predictor-outcome pairs $(\vec{x}, y)$, so we are protected against overfitting. Miller, Hui, and Tierney (1990) and Jakulin and Bratko (2003) discuss the use of scores to summarize validation performance in logistic regression.

Maximizing the Brier score [Brier (1950)] is equivalent to minimizing mean square error, and maximizing the logarithmic score is equivalent to maximizing the likelihood of the out-of-sample data. Both these rules are "proper" in the sense of being maximized by the true probability, if the model is indeed true [Winkler (1969)].

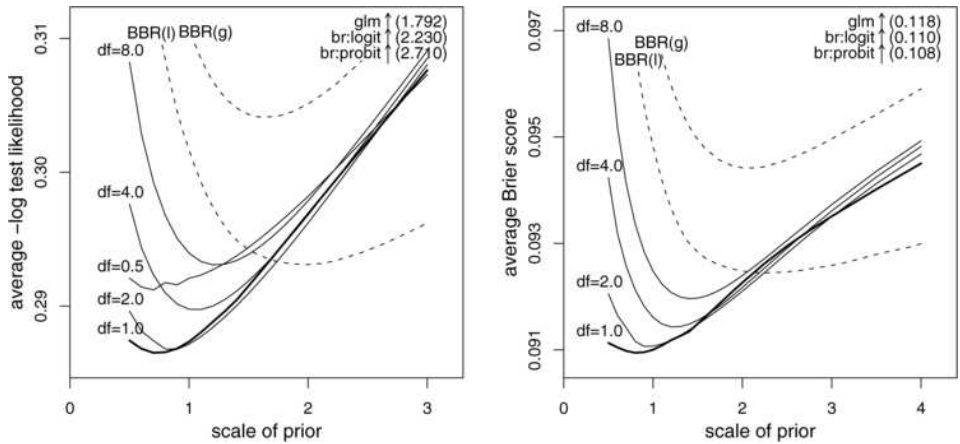

FIG. 6. *Mean logarithmic score (left plot) and Brier score (right plot), in fivefold cross–validation averaging over the data sets in the UCI corpus, for different independent prior distributions for logistic regression coefficients. Higher value on the y axis indicates a larger error. Each line represents a different degrees-of-freedom parameter for the Student-t prior family. BBR(l) indicates the Laplace prior with the BBR algorithm of Genkin, Lewis, and Madigan ([2007](#)), and BBR(g) represents the Gaussian prior. The Cauchy prior distribution with scale 0.75 performs best, while the performances of glm and brglm (shown in the upper-right corner) are so bad that we could not capture them on our scale. The scale axis corresponds to the square root of variance for the normal and the Laplace distributions.*

traditional default choice of maximum likelihood. True logistic regression coefficients are almost always quite a bit less than 5 (if predictors have been standardized), and so this Cauchy distribution actually contains less prior information than we really have. From this perspective, the uniform prior distribution is the most conservative, but sometimes too much so (in particular, for datasets that feature separation, coefficients have maximum likelihood estimates of infinity), and this new prior distribution is still somewhat conservative, thus defensible to statisticians. Any particular choice of prior distribution is arbitrary; we have motivated ours based on the notion that extremely large coefficients are unlikely, and as a longer-tailed version of the model corresponding to one-half success and one-half failure, as discussed in Section 2.2.

The BBR procedure of Genkin, Lewis, and Madigan [adapted from the regularization algorithm of Zhang and Oles ([2001](#))] employs a heuristic for determining the scale of the prior: the scale corresponds to $k/\mathrm{E}[\vec{x}\dot{\vec{x}}]$, where $k$ is the number of dimensions in $\vec{x}$. This heuristic assures some invariance with respect to the scaling of the input data. All the predictors in our experiments took either the value of 0 or of 1, and we did not perform additional scaling. The average value of the heuristic across the datasets was approximately 2.0, close to the optimum. However, the heuristic scale for individual datasets resulted in worse performance than using the corpus optimum. We interpret this observation as supporting our corpus-based approach for determining the parameters of the prior.

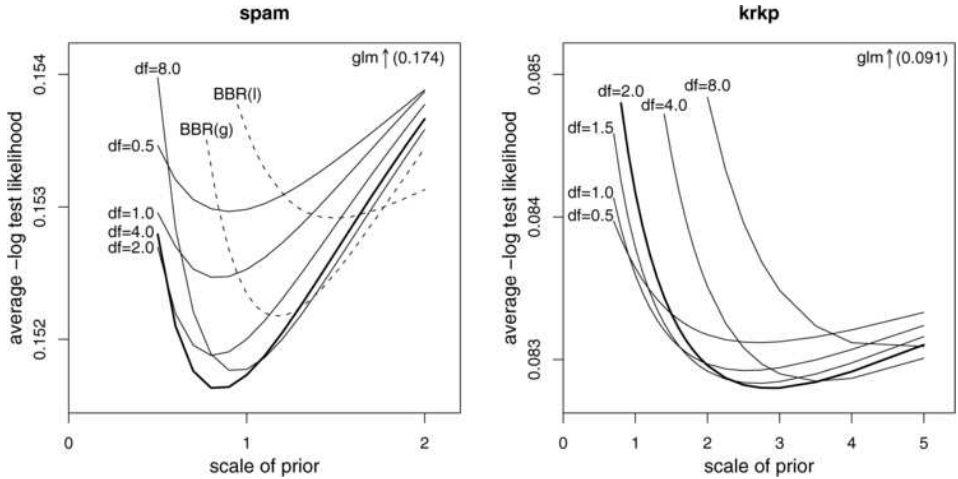

FIG. 7. *Mean logarithmic score for two datasets, "Spam" and "KRKP," from the UCI database. The curves show average cross-validated log-likelihood for estimates based on t prior distributions with different degrees of freedom and different scales. For the "spam" data, the $t_4$ with scale 0.8 is optimal, whereas for the "krkp" data, the $t_2$ with scale 2.8 performs best under cross-validation.*

**6. Discussion.** We recommend using, as a default prior model, independent Cauchy distributions on all logistic regression coefficients, each centered at 0 and with scale parameter 10 for the constant term and 2.5 for all other coefficients. Before fitting this model, we center each binary input to have mean 0 and rescale each numeric input to have mean 0 and standard deviation 0.5. When applying this procedure to classical logistic regression, we fit the model using an adaptation of the standard iteratively weighted least squares computation, using the posterior mode as a point estimate and the curvature of the log-posterior density to get standard errors. More generally, the prior distribution can be used as part of a fully Bayesian computation in more complex settings such as hierarchical models.

A theoretical concern with our method is that the prior distribution is defined on centered and scaled input variables, thus, it implicitly depends on the data. As more data arrive, the linear transformations used in the centering and scaling will change, thus changing the implied prior distribution as defined on the original scale of the data. A natural extension here would be to formally make the procedure hierarchical, for example, defining the $j$th input variable $x_{ij}$ as having a population mean $\mu_j$ and standard deviation $\sigma_j$, then defining the prior distributions for the corresponding predictors in terms of scaled inputs of the form $z_{ij} = (x_{ij} - \mu_j)/(2\sigma_j)$. We did not go this route, however, because modeling all the input variables corresponds to a potentially immense effort which is contrary to the spirit of this method, which is to be a quick automatic solution. In practice, we do not see the dependence of our prior distribution on data as a major concern, although we imagine it could cause difficulties when sample sizes are very small.

Modeling the coefficient of a scaled variable is analogous to parameterizing a simple regression through the correlation, which depends on the distribution of $x$ as well as the regression of $y$ on $x$. Changing the values of $x$ can change the correlation, and thus the implicit prior distribution, even though the regression is not changing at all (assuming an underlying linear relationship). That said, this is the cost of having an informative prior distribution: some scale must be used, and the scale of the data seems like a reasonable default choice. No model can be universally applied: in many settings it will make more sense to use a more informative prior distribution based on subject-matter knowledge; in other cases, where parameters might plausibly take on any value, a noninformative prior distribution might be appropriate.

Finally, one might argue that the Bayesian procedure, by always giving an estimate, obscures nonidentifiability and could lead the user into a false sense of security. To this objection, we would reply [following Zorn (2005)] as follows: first, one is always free to also fit using maximum likelihood, and second, separation corresponds to information in the data, which is ignored if the offending predictor is removed and awkward to handle if it is included with an infinite coefficient (see, for example, the estimates for 1964 in the first column of Figure 2). Given that we do not expect to see effects as large as 10 on the logistic scale, it is appropriate to use this information. As we have seen in specific examples and also in the corpus of datasets, this weakly-informative prior distribution yields estimates that make more sense and perform better predictively, compared to maximum likelihood, which is still the standard approach for routine logistic regression in theoretical and applied statistics.

**Acknowledgments.** We thank Chuanhai Liu, David Dunson, Hal Stern, David van Dyk, and editors and referees for helpful comments, Peter Messeri for the HIV example, David Madigan for help with the BBR software, Ioannis Kosimidis for help with the brglm software, Masanao Yajima for help in developing `bayesglm`, and the National Science Foundation, National Institutes of Health, and Columbia University Applied Statistics Center for financial support.

A. Gelman  
Department of Statistics  
  and Department of Political Science  
Columbia University  
New York  
USA  
E-mail: gelman@stat.columbia.edu  
URL: www.stat.columbia.edu/~gelman  

A. Jakulin  
Department of Statistics  
Columbia University  
New York  
USA  

M. G. Pittau  
Department of Economics  
University of Rome  
Italy  

Y.-S. Su  
Department of Political Science  
City University of New York  
USA